\documentclass[aps,showpacs,prl,reprint,superscriptaddress]{revtex4-1}

\usepackage{graphicx}
\usepackage[gnuplot]{pstricks,xcolor}
\usepackage{subcaption}

\begin{document}
\title{Impact of point defects on the electronic structure of paramagnetic CrN}

\author{David Holec}
\affiliation{Department of Physical Metallurgy and Materials Testing, Montanuniversit\"at Leoben, A-8700 Leoben, Austria}
\affiliation{Christian Doppler Laboratory for Application Oriented Coating Development, Institute of Materials Science
and Technology, Vienna University of Technology, A-1040 Vienna, Austria}
\email[]{david.holec@unileoben.ac.at}
\author{Liangcai Zhou}
\affiliation{Institute of Materials Science and Technology, Vienna University of Technology, Vienna, A-1140, Austria}
\author{Zaoli Zhang}
\affiliation{Erich Schmid Institute of Materials Science, Austrian Academy of Sciences, A-8700 Leoben, Austria}
\author{Paul H. Mayrhofer}
\affiliation{Christian Doppler Laboratory for Application Oriented Coating Development, Institute of Materials Science
and Technology, Vienna University of Technology, A-1040 Vienna, Austria}
\affiliation{Institute of Materials Science and Technology, Vienna University of Technology, Vienna, A-1140, Austria}

\newcommand{\uu}[1]{\,\mathrm{#1}}
\newcommand{\Et}{\ensuremath{E_{\mathrm{tot}}}}

\date{\today}

\begin{abstract}
This paper presents first principles calculations of paramagnetic cubic CrN$_x$ with the aim to provide a deeper insight into the effect of point defects on the electronic structure, and to make a direct comparison with a transmission electron microscopy-based study on this material. Among several types of point defects which may result in N-deficient material, N vacancy is found to be energetically preferred to Cr interstitial and anti-sites. Electron Energy Loss Near Edge Structure of N K-edge transition is calculated for various concentrations of N vacancies in CrN$_x$, yielding the same trends as experimentally observed. Analysis of the electronic structure reveals decreased charge transfer from Cr sites with increased N vacancy content, hence increasing the metallic character of the defected material. Finally, the electronic structure is found to be strongly dependent on the local environment (i.e. presence of the N vacancies).
\end{abstract}


\maketitle 

\section{Introduction}
CrN is an important material system used for hard protective coatings \cite{Navinsek1997-sg,Hultman2000-be,Mayrhofer2001-oq}. The CrN $0\uu{K}$ ground state is an orthorhombic structure with an anti-ferromagnetic (afm) ordering realised by (110) planes of alternating Cr spins up and down \cite{Corliss1960-oc,Miao2005-kx,Alling2010-to}. At a temperature $\approx 273\uu{K}$ \cite{Navinsek1997-sg}, a magneto-structural phase transition takes place, and CrN adopts cubic B1 (NaCl prototype) structure with paramagnetic (pm) state.

An extensive \textit{ab initio} work on transition metal nitrides in general, and on CrN-related materials in particular, providing insight and theoretical guidance for experiments has been performed in the past years \cite{Harrison1996-xa,Alling2007-tc,Alling2008-of,Rivadulla2009-at,Holec2010-mo, Rovere2010-fs,Alling2010-to,Alling2010-ju,Alling2010-mq,He2011-dg,Steneteg2012-us,Botana2012-te,Botana2013-uc,Alling2013-uv,Zhou2013-ze,Chawla2013-bx, Zhou2013-zx,Shulumba2014-ey,Zhou2014-wc}. The early works ignored the magnetic nature of pm-CrN \cite{Harrison1996-xa,Rivadulla2009-at,Zhang2013-oh} and simply treated it as non-magnetic material (i.e. fulfilling the condition of the macroscopic magnetic moment to be 0); however, this has been shown to lead to a significant overestimation of its bulk modulus \cite{Alling2010-ju} and an underestimation of lattice parameters \cite{Alling2010-ju,Zhou2014-wc}. Therefore, the most recent works considered explicitly the non-zero local magnetic moments. Nevertheless, the paramagnetic phase is challenging for description within the periodic crystalline model conveniently adopted by first principles calculations. Several methods have been proposed over the years \cite{Steneteg2012-us, Kormann2012-we, Shulumba2014-ey, Zhou2013-ze, Zhou2014-wc}. Here, we adopt a supercell-based approach \cite{Rovere2010-fs, Zhou2013-ze, Zhou2014-wc} with spins up and down being distributed on the Cr atoms according to a Special Quasi-random Structures (SQS) \cite{Wei1990-zt,Holec2013-yt} scheme, which provides a mathematically rigorous recipe for generating supercells with as random as possible arrangement of atoms (or spins in this particular case).

The electronic structure of perfect CrN has been addressed previously by \textit{ab initio} calculation \cite{Harrison1996-xa, Alling2008-of, He2011-dg, Botana2012-te, Botana2013-uc} as well as by experiments \cite{Zhang2010-qb,Zhang2011-bc, Zhang2013-oh}. In this paper we report on \textit{ab initio} calculations of stability changes induced by point defects in pm-CrN. The electronic structure and Electron Energy Loss Near Edge Structure (ELNES) spectra are discussed in conjunction with N vacancies, and the results are carefully compared with recent experimental reports \cite{Zhang2011-bc, Zhang2013-oh}.

\section{Calculational details}

The structural models for cubic pm-CrN including vacancies were generated according to the SQS method \cite{Wei1990-zt, Zhou2013-zx,Zhou2014-wc} by optimising the short-range order parameters up the 5\textsuperscript{th} nearest neighbour distance for supercells containing 64 sites (32 Cr and 32 N). The paramagnetic state was modelled by assuming Cr spin up and Cr spin down as two inequivalent species within the SQS scheme. The supercells were structurally relaxed using the Density Functional Theory \cite{Kohn1965-rd,Hohenberg1964-in} implemented in Vienna Ab initio Simulation Package \cite{Kresse1996-tg,Kresse1996-gt} employing projector augmented wave pseudopotentials \cite{Kresse1999-if} with the local (spin) density approximation (L(S)DA) \cite{Kohn1965-rd} for the exchange and correlation effects. Additionally, the on-site Coulomb interaction term, $+U=3\uu{eV}$, was applied to Cr $d$-states to improve the electronic structure description  \cite{Alling2010-to,Alling2010-ju,Steneteg2012-us,Shulumba2014-ey}. Plane-wave cut-off energy was set to $500\uu{eV}$ and the Monkhorst-Pack mesh had $6\times6\times6$ $k$-points.

\begin{figure*}[t!]
  \begin{subfigure}[t]{0.45\textwidth}
    \caption{}\label{fig:Ef}
    \includegraphics[height=0.7\columnwidth]{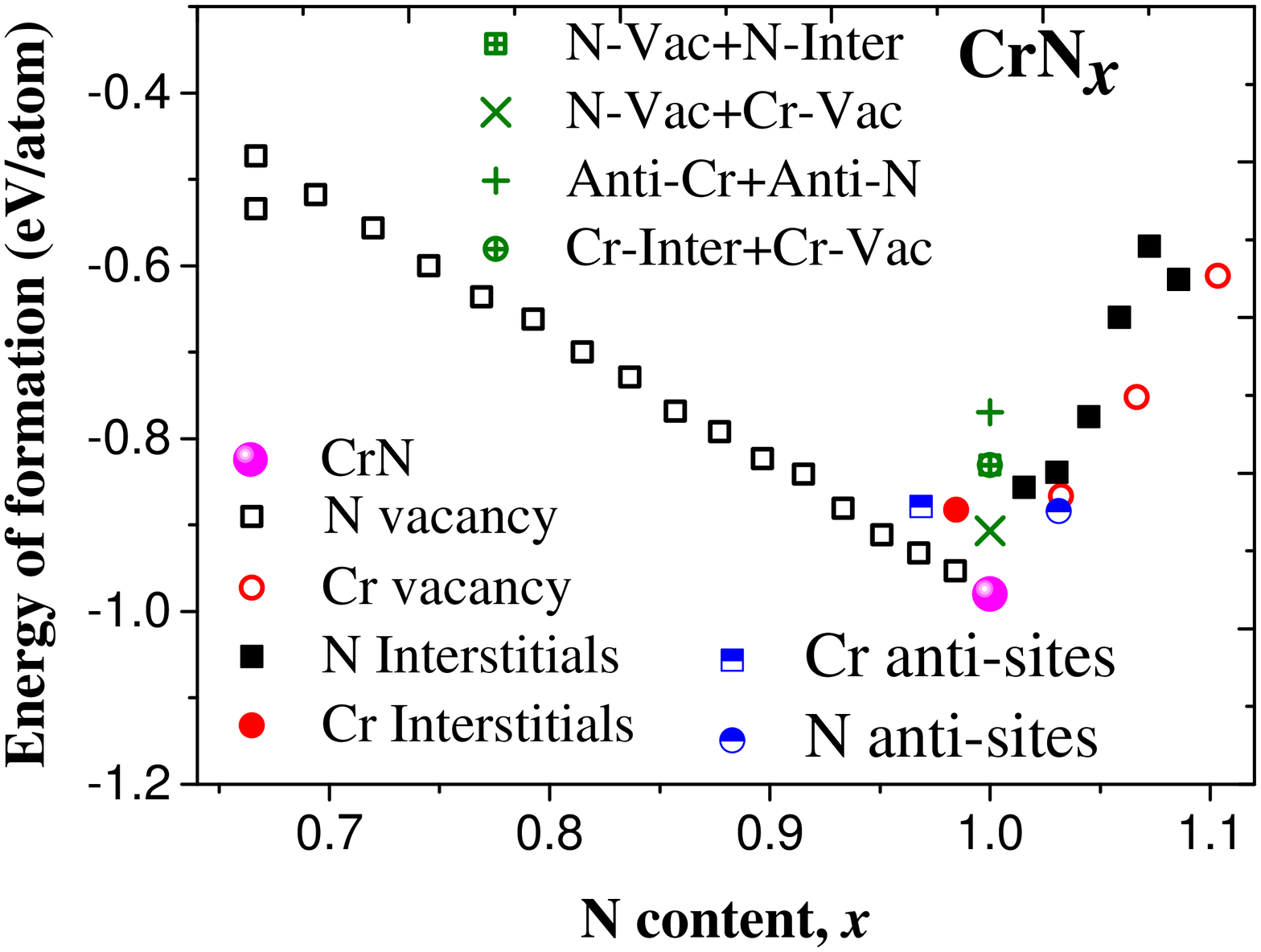}
  \end{subfigure}
  \begin{subfigure}[t]{0.45\textwidth}
    \caption{}\label{fig:aLat}
    \includegraphics[height=0.7\columnwidth]{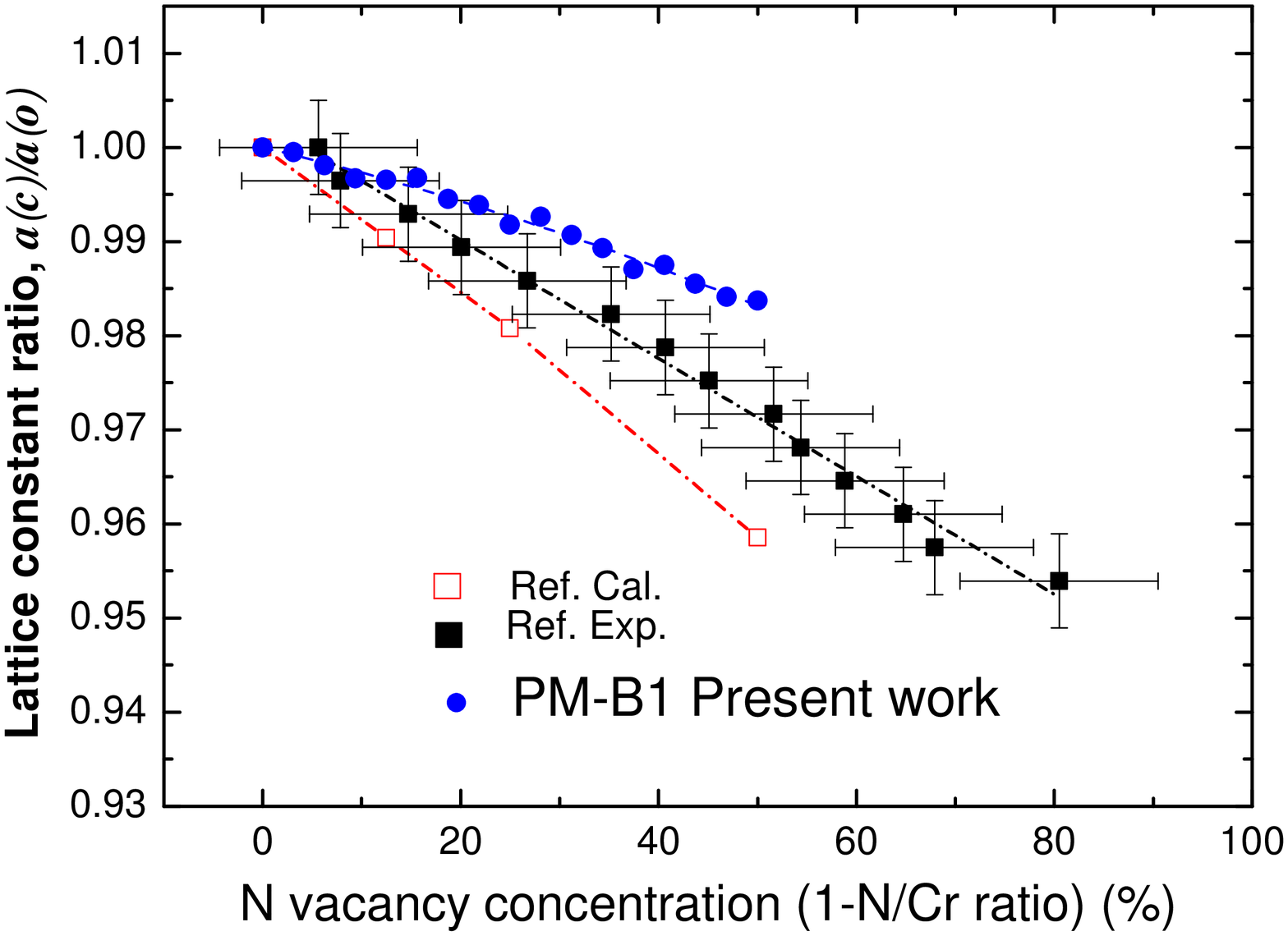}
  \end{subfigure}
  \caption{(a) Energy of formation, $E_f$, as a function of N content in CrN$_x$ with various point-defects described in the text. (b) Cubic lattice parameter, $a$, of N deficient CrN$_x$ relative to the value $a_0$ of perfect stoichiometric CrN. Experimental and non-magnetic calculated values are taken from Ref.~\cite{Zhang2013-oh}.}
\end{figure*}

The Electron Energy Loss Near Edge Structures (ELNES) were modelled using a Telnes program which is a part of the Wien2k, an all-electron full-potential implementation of DFT. The optimised supercells from VASP were taken as inputs for the structural description of paramagnetic CrN$_x$. Spherical harmonics up to $l_{\max}=10$ and plane-waves up to a cut-off defined by $R_{MT}K_{\max}=7$ together with the muffin-tin radii of $R_{MT}(\text{Cr})=2.00\uu{Bohr}$ and $R_{MT}(\text{N})=1.72\uu{Bohr}$) were used for expansion of the total wave function on an automatically generated grid with 45 $k$-points. The ELNES were calculated using a Slater's transition state in which $0.5\uu{e}$ from the initial core state was put into the conduction band (final states) \cite{Holec2011-xb}.

\section{Results and discussion}

\subsection{Stability of point defects in CrN}

Energy of formation, $E_f$, is a measure of the chemical stability of a compound. It is calculated as
\begin{equation}
  E_f(\mathrm{CrN}_x)=\Et(\mathrm{CrN}_x)-\frac{M\Et(\mathrm{Cr}^{\mathrm{bcc}})+N/2\Et(\mathrm{N}_2)}{M+N}
\end{equation}
where $\Et(X)$ is the total energy of $X$ per atom in its stable configuration, and $M$ and $N$ are the numbers of Cr and N atoms, respectively, in the (defected) supercell representing CrN$_x$.

Our calculations show that out of all the here investigated defects, the most stable configuration is the perfect (undefected) CrN (see Fig.~\ref{fig:Ef}). Both N vacancies and Cr vacancies yield almost linear increase of the energy of formation as their concentration increases. The overall N content can be decreased by N vacancies as well as by an introduction of Cr interstitial or Cr anti-sites. However, our predictions suggest that the latter two types of defects possess significantly higher energy than structures with the same chemistry obtained using N vacancies. We therefore conclude that when CrN material is N-deficient, it happens most likely due to the formation of N vacancies, which is in agreement with experimental observations \cite{Zhang2013-oh}.

The situation is not straightforward for the N-rich side ($x>1$) of the diagram in Fig.~\ref{fig:Ef}. For example, for $x\approx1.04$, $E_f$ of structures with N interstitials is only $28\uu{meV/at.}$ higher than when the same composition is obtained by Cr vacancies. Interestingly enough, the overall lowest energy exhibits a configuration with N anti-sites, which is $\approx17\uu{meV/at.}$ more favourable than the Cr vacancy configuration. Nevertheless, since this differences are in the range of thermal vibrations at room temperature $k_BT\approx25\uu{meV}$, we predict that Cr vacancies in addition to N anti-sites are to be expected in experimental samples with N-rich compositions at room temperature.

Finally we note that for stoichiometric CrN our calculations clearly favour a perfect crystal to structures containing Frenkel defects (an interstitial--vacancy pair), vacancies or anti-sites.

The predicted cubic lattice parameter of the pm-CrN as a function of the N content is shown in Fig.~\ref{fig:aLat} together with experimental and calculated data from Ref.~\cite{Zhang2013-oh}. Our predictions exhibit slightly slower decrease of the lattice parameter with decreasing N content as compared with the experimental data, while the previously published calculated lattice parameter shows more rapid decrease. We ascribe this difference to the fact that the calculations in Ref.~\cite{Zhang2013-oh} were performed on rather small supercells (16 atoms as compared with 64 used in the present work) and neglecting the magnetic effects. The discrepancy between the here predicted and experimental slope can be rationalised by the fact that the vacancies are randomly distributed in our structural models while experimentally they were reported to be ordered \cite{Zhang2013-oh}.

\begin{figure}
  \centering
  \includegraphics[width=8cm]{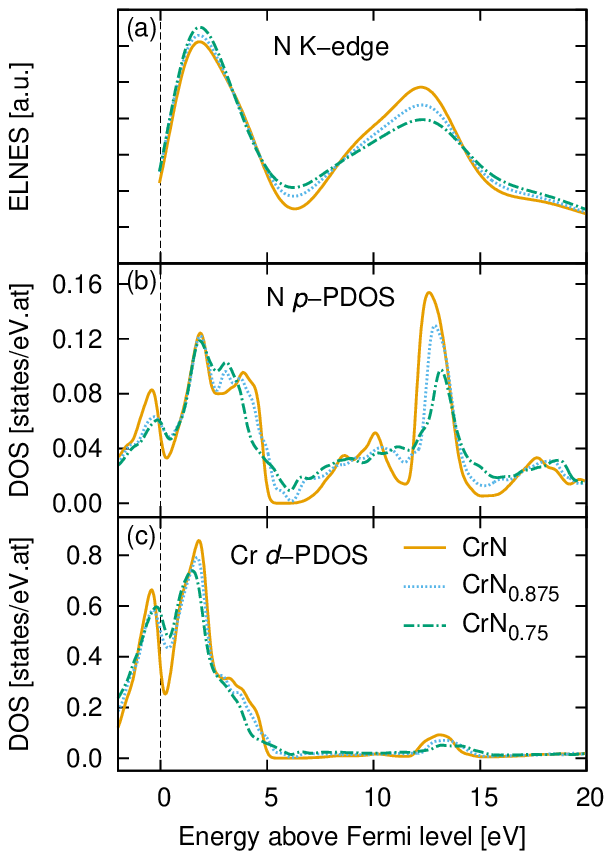}
  \caption{(a) N K-edge ELNES, (b) N $p$-PDOS, and (c) Cr $d$-PDOS for pm-CrN (solid line), pm-CrN$_{0.875}$ (dotted line), and pm-CrN$_{0.75}$ (dash-dotted line).}
  \label{fig:ELNES}
\end{figure}

The calculated evolution of the N K-edge (N 1$s\to$2$p$ transition) ELNES with increasing amount of N vacancies is shown in Fig.~\ref{fig:ELNES}a. The spectra are averaged over all N sites in the supercell, and are normalised to the intensity of the first peak at about $2\uu{eV}$. The spectra of the cubic phase exhibit also a second peak at about $12\uu{eV}$, whose intensity gradually decreases with increasing amount of the nitrogen vacancies. This is qualitatively the same behaviour as observed in experiments \cite{Zhang2013-oh}.

A comparison of the calculated N K-edge with the density of states projected on N $p$- and Cr $d$-states revels two facts. Firstly, N K-edge corresponds well with the (strongly) broadened N $p$-PDOS (similarly to e.g., previous reports for Ti$_{1-x}$Al$_x$N \cite{Holec2011-xb}). This is, however, expected as the unoccupied N $p$-states are the final states of the N K-edge excitation. The small peak at around $10\uu{eV}$ above the Fermi level is responsible for the shoulder at the lower-energy side of the second N K-edge peak, hence causing its asymmetry. Secondly, the unoccupied N $p$-states are well hybridised with the Cr $d$-states, which is related to formation of bonding and anti-bonding $sp^3d^2$ orbitals (the latter in the conduction band). Hence, when N vacancies are introduced, this hybridisation becomes weaker which is reflected by broadening of the peak at around $10\uu{eV}$ above the Fermi level together with lowering the peak intensity of the second PDOS peak at $\approx13\uu{eV}$ above Fermi level. Consequently, the amount of states in the region of the second N K-edge ELNES peak becomes smaller with increasing N vacancy content, causing the corresponding intensity drop.

\begin{figure*}[ht]
  \centering
  \begin{subfigure}[m]{7cm}
    \caption{}\label{fig:loc_env_curves}\medskip
   \includegraphics[width=7cm]{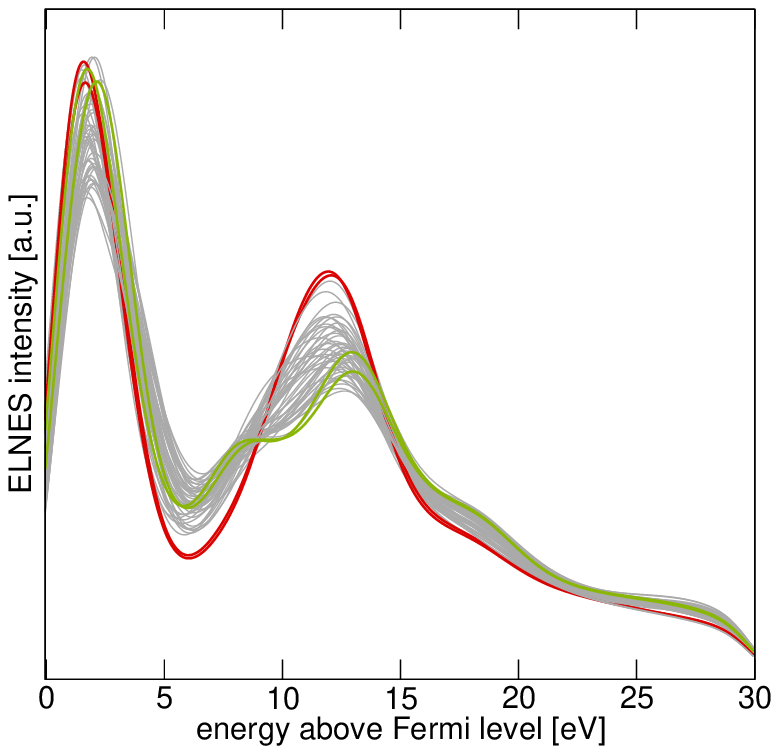}
  \end{subfigure}\hspace*{1cm}
  \begin{subfigure}[m]{7cm}
    \caption{}\label{fig:loc_env_atoms}\medskip
    \pspicture(7cm,7cm)
      \rput(3.5cm,4cm){\includegraphics[width=6cm]{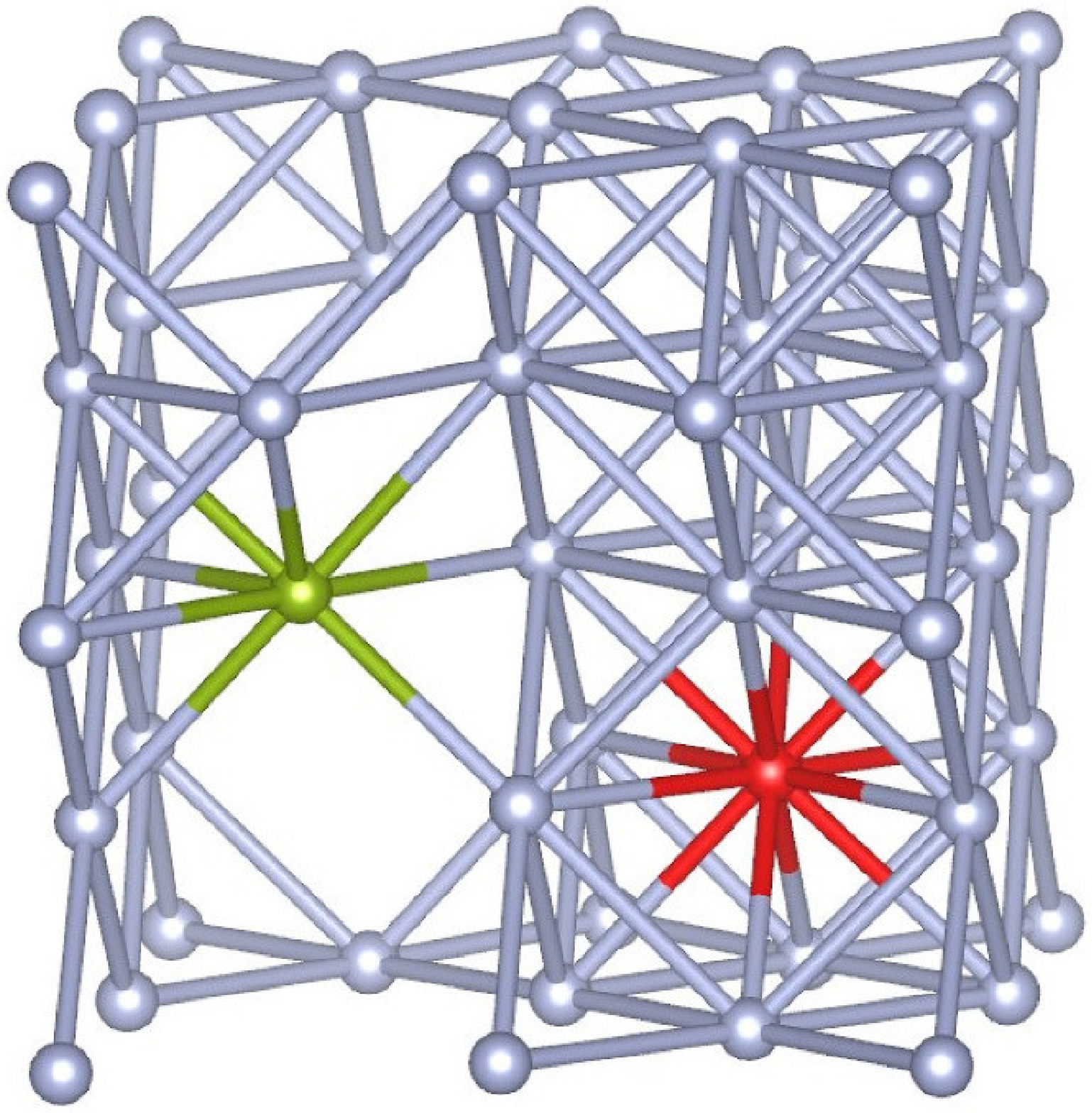}}
      \rput(-8cm,0){\psline[linewidth=1.5pt,arrows=->,arrowsize=2.5mm,arrowinset=0,linestyle=dashed](4.4cm,2.2cm)(9.8cm,3.6cm)}
      \rput(-8cm,0){\psline[linewidth=1.5pt,arrows=->,arrowsize=2.5mm,arrowinset=0](3.2cm,4cm)(12.4cm,2.8cm)}
    \endpspicture   
  \end{subfigure}
  \caption{(a) Impact of the local environment on the calculated N K-edge ELNES. (b) N sublattice with vacancies. The coloured sites in (b) correspond with the same coloured curves in the graph in (a), representing ideally stoichiometric (red, solid line arrow) and strongly N-deficient (green, dashed arrow) environments.}
  \label{fig:loc_env}
\end{figure*}

The curves in Fig.~\ref{fig:ELNES} represent macroscopic trends since they were obtained by averaging the ELNES or PDOS curves over all N or Cr sites. However, the same trends as described above are present also in a single defected cell in which N and Cr atoms with different numbers of the second and the first nearest neighbours, respectively, exist due to the presence of vacancies. This is clearly demonstrated in Fig.~\ref{fig:loc_env} showing that when the excitation takes place on a N atom with a bulk-like environment, the N K-edge ELNES resembles that of perfect CrN, while N site with a high amount of N vacancies in its neighbourhood results in splitting of the second peak (compare Figs.~\ref{fig:loc_env_curves} and \ref{fig:loc_env_atoms}).

The N K-edge onset could be estimated as a difference in the total energy of an unperturbed (initial state) and excited (final state with a full core hole) system \cite{Holec2011-xb}. This analysis yields values of $395.3$, $400.9$, and $401.1\uu{eV}$ for CrN, CrN$_{0.875}$, and CrN$_{0.75}$, respectively. Although the absolute values do not agree with the data from Ref.~\cite{Zhang2013-oh}, the trend that the edge onset increases with the N vacancy concentration is well reproduced. The same procedure applied to CrN$_2$ yields a valued of $401.7\uu{eV}$ for the N K-edge onset. Again, the qualitative fact that N K-edge onset increases from CrN to CrN$_2$ agrees well with the observations reported in \cite{Zhang2011-bc}.

\begin{figure}[ht]
  \centering
  \includegraphics[width=8cm]{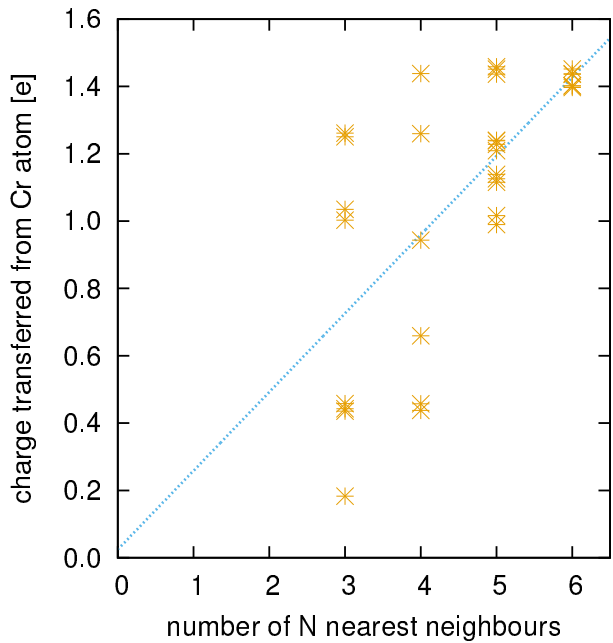}
  \caption{Bader's analysis of the charge transferred from Cr sites in CrN$_{0.75}$ supercell. The dotted line represents a linear fit to the data ($R^2=0.44$).}
  \label{fig:bader}
\end{figure}

In Ref.~\cite{Zhang2013-oh} it is argued that the increased number of N vacancies leads to an increased metallic character of the compound. We have performed Bader's analysis \cite{Bader1990-kt} to quantify the charge transfer from Cr to N atoms. In perfect CrN, approximately $1.45\uu{e/at.}$ is transferred from Cr atoms (hence making them positively charged cations) to N atoms (thus becoming negatively charged anions). When performing the same analysis in a supercell with 25\% of N sites being vacant, the average extra charge on remaining 75\% N sites is $\approx1.43\uu{e}$, i.e. it remains almost unchanged with respect to the N sites in CrN. For the cell to be neutral this means that each Cr site looses on average a smaller fraction of its charge ($\approx1.08\uu{e}$), consequently leaving more charge ``available'' to participate in metallic boding. The strong dependence of the charge transfer on the local environment is presented in Fig.~\ref{fig:bader}. Although the data are somewhat scattered, a trend that the charge transferred from a Cr site is proportional to the number of N nearest neighbours is envisioned (see the dotted linear fit in Fig.~\ref{fig:bader}). The larger scatter for the Cr sites next to vacancies suggests that also the longer range interactions are not negligible. We therefore conclude that it is directly the vacancy-altered charge transfer which yields reduced ionic and profound metallic bonding character rather than changed lattice constant reduction as speculated in Ref.~\cite{Zhang2013-oh}.

To strengthen our conclusions, we have performed a number of additional calculations (not shown here): The N K-edges are almost identical irrespective of whether L(S)DA or generalised gradient approximation (GGA) \cite{Perdew1996-vd} is used for the exchange-correlation effects. On the other hand, to neglect the magnetic nature of CrN seems to be much more serious: each of the two main N K-edge peaks splits into two when CrN is treated non-magnetically. For the ferromagnetic arrangement of spins, the resulting edge shape is very close to the paramagnetic-state curve.

\section{Conclusions}

We have performed an \textit{ab initio} study on stability of point defects in CrN and the electronic structure changes induced by N vacancies. Our calculations of energy of formation and lattice parameters confirmed that the experimentally observed N understoichiometry is likely to be related to N vacancies rather than to any other point defect. Simulations of a N K-edge ELNES evolution with amount of N vacancies yields the same behaviour as observed previously by transmission electron microscopy: The edge shape has two peaks separated by $\approx10\uu{eV}$. With the increasing content of N vacancies the relative intensity of the second peak (with respect to the first peak) decreases, and a shoulder at its lower-energy side increases. We ascribe these changes to the decreased hybridisation of N $p$- and Cr $d$-states, and a decreased ionicity as more N vacancies are present. Our results demonstrate the strong dependence of ELNES on the local environment of a site where the excitation takes place.

\section*{Acknowledgements}
The financial support by the START Program (Y371) of the Austrian Science Fund (FWF), as well as by the Austrian Federal Ministry of Economy, Family and Youth and the National Foundation of Research, Technology and Development is greatly acknowledged. The calculations were partly performed using the CPU time at the Vienna Scientific Cluster (VSC).

\bibliography{refs}
\end{document}